\documentclass[aps,pre,preprintnumbers,amsmath,twocolumn,amssymb,showpacs]{revtex4}
\usepackage{graphics}
\usepackage{bm}
\usepackage{dcolumn}
\usepackage{epsfig}
\usepackage{subfigure}
\usepackage{lipsum}
\usepackage{amsmath}
\usepackage{mathtools}

\def\noi{\noindent}
\def\bc{\begin{center}}
\def\ec{\end{center}}
 \newcommand{\bea}{\begin{equation}}
 \newcommand{\eea}{\end{equation}\noi}
 \newcommand{\ber}{\begin{eqnarray}}
 \newcommand{\eer}{\end{eqnarray}\noi}
\begin{document}
\title{Universal spatio-temporal scaling of distortions in a drifting lattice}
\author{Pritha Dolai$^1$}
\author{Abhik Basu$^2$}\email{abhik.basu@saha.ac.in} \author{Aditi 
Simha$^{1}$}\email{phyadt@iitm.ac.in}
\affiliation{$^1$Department of Physics, Indian Institute of Technology Madras, 
Chennai 600036, India\\$^2$Saha Institute of Nuclear Physics, 1/AF 
Bidhannagar, Calcutta  700064, India}


\begin{abstract}
We study the dynamical response to small distortions of a lattice about its uniform state, 
drifting through a dissipative medium due to an external force, and show, analytically 
and numerically, that the fluctuations, both transverse and longitudinal to the direction of 
the drift, exhibit spatiotemporal scaling belonging to the Kardar-Parisi-Zhang 
universality class. Further, we predict that a colloidal crystal drifting in a constant 
electric field is linearly stable against distortions and the distortions propagate as 
underdamped waves.
\end{abstract}
\maketitle
\section{Introduction}
It is well known from elastic theory that distortions in a crystal at thermal equilibrium propagate as 
waves with a speed determined by the elastic constants of the lattice \cite{elastic,martin}. 
The response of a lattice drifting due to an external force through a dissipative 
medium was first addressed by Lahiri and Ramaswamy (LR) in ~\cite{Lahiri}. 
The linear stability of the lattice was predicted to depend on certain model parameters that 
govern the strain-dependence of the mobility of the lattice.
The role of 
anharmonic effects and random fluctuations (possibly of nonequilibrium origin) on the 
macrocscopic nature of steady states, including scaling
properties is still unknown. This potentially opens up the
possibility that either the anharmonic effects drive the ensuing
steady state away from its equilibrium counter part, or 
leave the system macroscopically indistinguishable from a crystal in equilibrium.
In this letter, we address these issues. Specifically, we ask: what is the macroscopic nature of 
the drifting non-equilibrium state? 



The study of drifting lattices began with the work of Crowley in 1971 ~\cite{Crowley} who predicted that an array
of particles moving through a viscous fluid is unstable to clumping due to hydrodynamic forces 
alone, a result he verified experimentally by dropping steel balls into turpentine oil.  
The role of elastic and Brownian forces on this lattice instability was analysed by Lahiri and Ramaswamy in 1997
 ~\cite{Lahiri}\,. A set of continuum equations for the displacement fields of the drifting lattice, constructed
 using symmetry agruments, showed that the lattice was linearly unstable to clumping,  
even in the presence of elasticity. The role of nonlinearities and 
noise on the linear instability was not analysed. Numerical studies of an equivalent lattice-gas model 
describing the coupled
dynamics of concentration and tilt fields showed that the lattice was stable to distortions upto a critical
P\'{e}clet number at which a nonequilibrium phase transition to a clumped state occured. 

In this work, we find that the nonequilibrium steady state of the drifting lattice is phenomenally
different from its equilibrium counterpart. We show that small, 
long wavelength lattice distortions, exhibit spatiotemporal scaling both 
transverse and longitudinal to the direction of drift of the lattice  
and establish, analytically and numerically, that the fluctuations display dynamical 
scaling that belongs to the Kardar-Parisi-Zhang (KPZ) universality class~\cite{kpz}. 
As an example of this drifting nonequilibrium state, we analyse the dynamics
of distortions in a colloidal crystal drifting in a constant electric field and show that 
it has a linearly stable state in which long wavelength distortions propagate as under 
damped waves. The wave speeds and the length scale beyond which these propagating waves can 
be detected are also calculated in terms of the driving force and the parameters defining 
their interactions.


\section{Drifting lattices in disspative media}
For a driven, nonequilibrium system such as ours, the equations of motion 
for the degrees of freedom must be written down directly, by using 
symmetry arguments.
Physically, the equations of motion for the displacement field $\mathbf{u}(\mathbf{r},t)$ of a lattice 
moving in a frictional medium, ignoring inertia completely, must obey the equation
\begin{equation}
 \dot{\mathbf{u}}=\underline{\underline{\mathbf{M}}}(\nabla \mathbf{u}).\,\mathbf{F}_{tot} = 
 \underline{\underline{\mathbf{M}}}(\nabla \mathbf{u}).(\mathbf{F}+ 
 \underline{\underline{\mathbf{D}}} \nabla \nabla \mathbf{u}+\boldsymbol{\eta})\, . \label{force}
\end{equation}
Here, $\underline{\underline{\mathbf{M}}}$ is the mobility 
tensor that depends on the local lattice strain, $\mathbf{F}_{tot}$ is the total force consisting of 
the external driving force $\mathbf{F}$, elastic forces due to lattice distortions 
$\underline{\underline{\mathbf{D}}}\nabla \nabla \mathbf{u}$  and the random force 
$\boldsymbol{\eta}$ acting on the particle due to the surrounding fluid. The mobility 
tensor has the form 
$\underline{\underline{\mathbf{M}}}=\underline{\underline{\mathbf{M_{0}}}}+\underline{\underline{\mathbf{A}}}
(\nabla \mathbf{u})+O(\nabla \mathbf{u})^{2})$ where 
$\underline{\underline{\mathbf{M_{0}}}}$ is the mean mobility of the undistorted lattice,  
$\underline{\underline{\mathbf{A}}}$ is the first order correction to it due to lattice distortions 
and the successive terms higher order corrections ~\cite{Lahiri}. These terms arise from 
interactions between particles in the surrounding viscous medium. 

For a lattice in the $(x,y)$ plane drifting along the $\hat{z}$ direction 
the equations of motion for the displacement field $(\mathbf{u}_{\perp},u_z)$ are isotropic in 
the transverse ($\perp$ or $(x,y)$\,) plane but not invariant under $z \rightarrow -z$. 
The equations hence have the form: 
\begin{eqnarray}
\mathbf{\dot{u}}_{\perp}= \lambda_{1}\partial_{z}\mathbf{u}_{\perp}+ 
\lambda_{2}\boldsymbol{\nabla}_{\perp}u_{z}+
D_{1}\boldsymbol{\nabla}_{\perp}^{2}\mathbf{u}_{\perp}+
D_{3}\partial_{z}^{2}\,\mathbf{u}_{\perp}+ \nonumber \\ 
O(\nabla u \nabla u) +\boldsymbol{\eta}_{\perp},\label{dotuperp} \\
\dot{u}_{z}= \lambda_{3}\boldsymbol{\nabla}_{\perp}\cdot \mathbf{u}_{\perp}+ \lambda_{4}\partial_{z}{u}_{z}
+D_{2}\boldsymbol{\nabla}_{\perp}^{2}u_{z}+D_{4}\partial_{z}^{2}u_{z}+\nonumber \\
D_{5}\partial_{z}\boldsymbol{\nabla}_{\perp}\cdot \mathbf{u}_{\perp}+ O(\nabla u \nabla u) +\eta_{z}.\label{dotuz}
\end{eqnarray}
These follow from eq.(\ref{force}) albeit in the frame of the drifting lattice. The constant term 
in (\ref{force}) has hence been omitted.
The $\lambda_{i}\,$s are phenomenological parameters arising from the strain dependence of the mobility 
and depend crucially on the details of the hydrodynamic interaction 
between particles in the system. They are proportional to the drift speed 
of the lattice. The $D_{i}\,$s are diffusion constants coming from
elastic restoring forces in eq.~(\ref{force})\,. $\eta_{\perp}$ and $\eta_{z}$
are Gaussian white noise in the lattice plane and perpendicular to it respectively.
There are a total of nine quadratic nonlinearities, $O(\nabla u \nabla u)$ terms, in these equations 
which arise from the dependence of the $\lambda_i\,$s on the local 
concentration and tilt ($\boldsymbol{\nabla_{\perp}}u_z$). 

In this paper, we work with a simplified version of these equations in one dimension \cite{Lahiri}. The 
displacement field $\mathbf{u}(\mathbf{r},t)$  of the lattice then has only two components $(u_x,u_z)$ and 
only derivatives in $x$ are considered; those along the direction of drift $\hat{z}$ are averaged out.
With this simplification eqs.(\ref{dotuperp},\ref{dotuz}) reduce to  
\begin{eqnarray}
\dot{u}_{x}= \lambda_{2}\partial_{x}u_{z}+
\gamma_{1}\partial_{x}u_{x}\partial_{x}u_{z}+D_{1}
\partial_{x}^{2}u_{x}+\eta_{x}\,,\hspace{0.5cm}\label{dotux} \\
\dot{u}_{z}= \lambda_{3}\partial_{x}u_{x}+\gamma_{2}(\partial_{x}u_{x})^{2}+\gamma_{3}
(\partial_{x}u_{z})^{2}+D_{2}\partial_{x}^{2}u_{z}+\eta_{z} \label{dotuz2}
\end{eqnarray}
Only 3 quadratic nonlinearities are allowed by symmetry and only eq.(\ref{dotuz2}) 
has the KPZ nonlinearity  $(\partial_{x}u_{z})^{2}\,$. The equations are coupled at the linear level
and can be decoupled, at the linear level, for fields that 
are appropriate linear combinations of $u_{x}, u_{z}$ (see eqs. (\ref{phi+eq},\ref{phi-eq})).
Equations of this type have been studied extensively in recent years in various contexts
(\cite{Spohn1,Spohn1b,Beijeren,Spohn3,Spohn3b,Spohn4,Das}).

Linearising and Fourier transforming the above equations in space and time, as in \cite{Lahiri}, yields the 
dispersion relations for the two modes of the system -- 
\begin{equation}
\omega=\frac{-ik^{2}(D_{1}+D_{2})}{2}\pm \frac{1}{2}\sqrt{4\lambda_{2}\lambda_{3}k^{2}-k^{4}(D_{1}-D_{2})^{2}}
\end{equation}
$\omega$ is the frequency and $k$ the wave number of the mode. 
For long wavelength (small $k$) distortions this implies that the crystal 
is linearly stable only when $\lambda_{2}\lambda_{3}>0$.
Symmetry arguments alone cannot {\it apriori} determine whether the lattice is stable as the signs 
of these parameters depend on the details of the interaction between particles which is system dependent. 
For a sedimenting lattice the product $\lambda_{2}\lambda_{3}$ was calculated and found to be negative 
\cite{Lahiri} implying a linear instability towards clumping. We calculate $\lambda_{2}\lambda_{3}$ for a 
colloidal crystal drifting due to an applied electric field before we address the effect of nonlinearities.


\begin{figure}
\begin{center}
{ 
\includegraphics[scale=0.53,keepaspectratio=true]{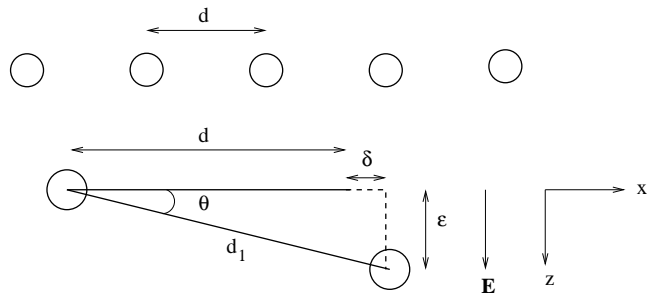}
}
\caption{Schematic diagram to show displacement fields. $\delta$ and $\epsilon$ 
are the displacements of a particle in $x$ and $z$ direction from its original position.}
\end{center}
\end{figure}
\section{Colloidal crystal in an electric field}
Consider a 1D lattice of colloidal particles of radius $a$ with lattice spacing $d$ 
in the $x$ direction and the electric field 
$\mathbf{E}$ perpendicular to the lattice (as in Fig.1). A single charged 
colloid drifts in the field with constant velocity, 
${\mathbf{V}}=\xi {\mathbf{E}}$, where $\xi$ is its 
mobility. Its motion results from a complex interplay of electrostatic, hydrodynamic and 
thermal forces and its mobility depends on various parameters such as the thickness of the 
electric double layer of small counterions, surface properties, charge density, ion 
concentration, and lipophilicity of the colloid and the specific
properties of counterions and salt ions. There is as yet 
no expression for the mobility applicable, in general, as a function of these parameters 
(\cite{Ohshima4,Ohshima5,Ohshima7,Lizana}). 
The mobility of a charged sphere, in the thin double layer limit, was first derived  
by Smoluchowski \cite{Smoluchowski} to be $\xi_{0}=\epsilon \zeta/\bar{\eta}$ where 
$\epsilon$ and $\bar{\eta}$ are the dielectric permittivity and viscosity of the 
colloidal solution and $\zeta$ the Zeta potential on the surface of the sphere. 
For double layers of arbitrary thickness but small $\zeta$, Smoluchowski's result for the 
mobility was modified by Henry to  $\xi=\xi_{0}f(\kappa a)$ where $\kappa^{-1}$ is the 
Debye length \cite{Henry} and $f(\kappa a)$ Henry\textquoteright{s} function which is an 
increasing function of $\kappa a$\,. The mobility of a particle
 is modified in the presence of other particles due to interactions between them. 
For two identical spherical particles of radius $a$, the mobility was derived 
using the method of reflections by Ennis {\it et al} ~\cite{Ennis}.
The electrophoretic velocity of a sphere in the presence of an 
identical sphere at a distance $d$ is given by 
\begin{equation}
\mathbf{V}=\frac{\xi}{4\pi}[A_{\lVert}\mathbf{ee}+A_{\bot}(\mathbf{I}-\mathbf{ee})+
 ( B_{\rVert}\mathbf{ee}+B_{\bot}(\mathbf{I}-\mathbf{ee}))].\mathbf{E} \label{vpair}
\end{equation}
Here $\mathbf{e}$ is a unit vector along the line joining the two spheres and 
$\mathbf{I}$ the unit tensor of rank two. $A_{||}$, $A_{\bot}$, $B_{||}$ and 
$B_{\bot}$
have the form (keeping only the leading order dependence on $d$\,):
$A_{||}=1-(\frac{a}{d})^{3}$\,, 
$A_{\bot}=1+\frac{1}{2}(\frac{a}{d})^{3}$\,,
 $B_{||}=(\frac{a}{d})^{3}\frac{L(\kappa a)}{f(\kappa a)}$ and 
 $B_{\bot}=-(\frac{a}{d})^{3}\frac{L(\kappa a)}{2 f(\kappa a)}$. 
The function $L(\kappa a)$ decreases monotonically with $\kappa a$\,.
The dominant interaction between two particles, as implied by this result, decays as $1/d^3$. 
Both $f(\kappa a)$ and $L(\kappa a)$ tend to 1 as $\kappa a \rightarrow \infty$. In this limit the 
result for thin diffuse layers is recovered where particles do not interact 
with each other. 

According to (\ref{vpair}) a pair of particles at distance $d$ apart (as in Fig.1) move in 
the $z$ direction with speed $v_0$ 
given by eq.(\ref{vpair})\,.
If one of them is displaced by $\delta$ and $\epsilon$ along and perpendicular, respectively, to 
the lattice at some instant of time, the change in velocity 
$\Delta v_{x}$ and $\Delta v_{z}$ along the $x$ and $z$ directions due 
to the displacement are
\begin{eqnarray}
\Delta v_{x}&=&C\left[\frac{3}{2}(\frac{\epsilon}{d})-6(\frac{\delta}{d})(\frac{\epsilon}{d})\right],
\label{dvx}\\
\Delta v_{z}&=&C\left[\frac{3}{2}(\frac{\delta}{d})-3(\frac{\delta}{d})^{2}+\frac{9}{4}(\frac{\epsilon}
{d})^{2}\right], \label{dvy}
\end{eqnarray}
where $C \approx v_0 (\frac{a}{d})^{3}(\frac{L(\kappa a)}{f(\kappa a)}-1)$. Using the expressions 
for $L(\kappa a)$ and $f(\kappa a)$ from \cite{Ennis} we find that $L(\kappa a)/f(\kappa a) > 1$, for all
$\kappa a$ and hence
$C>0$. This along with eqns.(\ref{dvy},\ref{dvx}) implies that the spheres fall slower when they are closer 
and a displacement 
along the field travels in the $+x$ direction. The implications of this for the drifting lattice are evident.
A perfect lattice drifts uniformly in the $z$ direction. If the lattice were perturbed, say a 
region of it compressed, then it would drift slower in this region. With time, this results in a tilt 
of the interfacial region between the compressed and uncompressed regions. These tilted regions drift
laterally as implied by eq.(\ref{dvx}). The direction of this lateral drift (given $C>0$) is such that
the tilted regions move apart dilating the compressed regions. The lattice is thus stable to distortions.
If we approximate $\partial_{x}u_{x} \approx \frac{\delta}{d}$ and  
$\partial_{x}u_{z}\approx\frac{\epsilon}{d}$, then
the expressions on the right hand side (RHS) of eqns.~(\ref{dvx},\ref{dvy}) are exactly the terms on the RHS of 
eqns.~(\ref{dotux},\ref{dotuz2}). The coefficients
$\lambda_{2}$, $\lambda_{3}$ for the drifting lattice can thus be obtained by summing the contributions 
of the nearest neighbors to the change in velocity 
$\Delta v_{x}$ and $\Delta v_{z}$ of a particle in the lattice. Our results for two particles allow us 
to conclude that $\lambda_{2} \lambda_{3} > 0$ since $C$ is always 
greater than zero. The speed of the propagating modes $v \propto \sqrt{\lambda_2 \lambda_3} \approx C$.
For particles of radius $a=1 \mu m$, $\kappa a= 2.5 $, $d=3a$ in an electric field of strength $150$ V$/m$, we 
estimate the speed of the propagating modes to be $~ 10 \mu m/s$.
These propagating modes dominate beyond a lengthscale 
$l_c\sim 2\pi D/\sqrt{\lambda_{2}\lambda_{3}}$. We estimate $l_c \approx 50\,d$ for this system.
It should hence be possible to detect these modes in systems that are larger than $l_c$.
A similar analysis for a 1D lattice drifting parallel to the electric field 
indicates that the lattice is linearly stable. This is a general result applicable to all drifting lattices.
Having established that the lattice is linearly stable, we ask what the effect of the noninearities and noise
are on this stable state. 

\section{Nonlinearities and fluctuations}
To analyze the effect of nonlinearities and fluctuations on the linearly stable state, approximate
methods must be used as eqns.~(\ref{dotux}-\ref{dotuz2}) cannot be solved in closed form. Exact results 
pertaining to their spatial and temporal scaling behavior can be obtained 
using a dynamic renormalization group (DRG) analysis~\cite{Stanley,halpin}. In particular, the
{\em roughness exponents} $\chi_x,\chi_z$ and {\em dynamic exponents} 
$z_x,z_z$ of the fields $u_x$ and $u_z$, respectively, defined by the scaling forms of their
correlation functions 
\begin{eqnarray}
C^{xx}(x,t)=\langle 
u_x(x,t)u_x(0,0)\rangle=A_x|x|^{2\chi_x}f_x(x/t^{z_x}),\label{corrux}\\
C^{zz}(x,t)=\langle 
u_z(x,t)u_z(0,0)\rangle=A_z|x|^{2\chi_z}f_z(x/t^{z_z}),\label{corruz}
\end{eqnarray}
can be determined using this method. Here the functions 
$f_x,f_z$ are dimensionless scaling functions of their arguments, and coefficients 
$A_x,A_z$ are constants.
On scaling space as $x\rightarrow bx$, time as $t\rightarrow b^{z_i}t$ and 
the fields as $u_i(x,t)\rightarrow 
b^{\chi_i}\,u_i(bx,b^{z_i} t)\,, i=x,z$, the correlation functions scale as $C^{xx}(x,t)\rightarrow 
b^{2\chi_x}C^{xx}(bx,b^{z_x}t),\,C^{zz}(x,t)\rightarrow b^{2\chi_z}C^{zz}(bx,b^{z_z}t)$. If 
$z_x=z_z$, then the model displays strong dynamic scaling, else  
weak dynamic scaling~\cite{Das}.

We begin by decoupling eqs.~(\ref{dotux},\ref{dotuz2}) at the linear level by defining
the fields $\phi_{\pm}=u_{x}\pm \nu u_{z} $ where $\nu=\sqrt{\lambda_{2}/\lambda_{3}}$\,.
In terms of $\phi_{\pm}$, they become 
\begin{eqnarray}
\dot{\phi}_{+}-\frac{\alpha}{2}\partial_{x}\phi_{+}+a_{1}(\partial_{x}\phi_{+})^
{2}+b_{1}(\partial_{x}\phi_{-})^{2} +\nonumber \\
c_{1}(\partial_{x}\phi_{+})(\partial_{x}\phi_{-})
= D_{+}\partial^{2}_{x}\phi_{+}+\eta_{+} \label{phi+eq}\,,\\
\dot{\phi}_{-}+\frac{\alpha}{2}\partial_{x}\phi_{-}+a_{2}(\partial_{x}\phi_{+})^
{2}+ 
b_{2}(\partial_{x}\phi_{-})^{2}+
\nonumber \\ c_{2}(\partial_{x}\phi_{+})(\partial_{x}\phi_{-})
= D_{-}\partial^{2}_{x}\phi_{-}+\eta_{-} \label{phi-eq}\,.
\end{eqnarray}
The co-efficient of the wave term 
$\frac{\alpha}{2}=\sqrt{\lambda_{2}\lambda_{3}}$, 
$\eta_{\pm}=\eta_{x}\pm \nu \eta_{z}$ and the 
co-efficients of nonlinear terms depend on $\lambda_{2}$, 
$\lambda_{3}$, $\gamma_{1}$, $\gamma_{2}$ and $\gamma_{3}$. 
The zero-mean Gaussian 
white noises $\eta_+,\eta_-$ are appropriate linear combinations of the noises 
$\eta_x,\eta_z$ and have correlations  $\langle\eta_{+}(x,t)\eta_{+}(x',t')\rangle = 
2A_{1}\delta(x-x')\delta(t-t')$ and 
$\langle\eta_{-}(x,t)\eta_{-}(x',t')\rangle = 2A_{2}\delta(x-x')\delta(t-t')\,$. 
$D_{+}$ and $D_{-}$ are the new diffusion constants. 
Noises $\eta_+$, $\eta_-$ have non-zero cross correlations of the form 
$\langle\eta_{+}(x,t)\eta_{-}(x',t')\rangle = 2A_{3}\delta(x-x')\delta(t-t')$.
Coupled equations of this type have been studied in considerable detail earlier.
We refer the reader to the work in  
(\cite{Spohn1,Spohn1b,Beijeren,Spohn3,Spohn3b,Spohn4}) for a perspective on this. Our approach
here is to use dynamic renormalisation group to extract the scaling properties of these 
equations. 
For the special case with $\gamma_{1}=2\gamma_{3}$ and $\gamma_{2}/\gamma_{3}=
\lambda_{3}/\lambda_{2}$ Eqs.~(\ref{phi+eq}-\ref{phi-eq}) reduce to two 
separate KPZ equations ~\cite{Lahiri}. 
\begin{figure}[t]
\begin{center}
{ 
\includegraphics[scale=0.27,keepaspectratio=true]{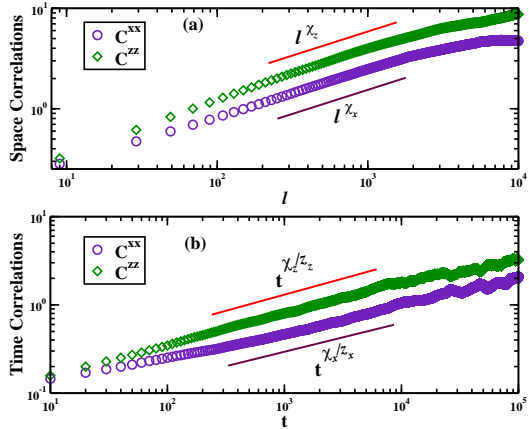}
}
\caption{(color online). Log-log plot of equal-time correlators $C^{xx}(l,t=0)$ and 
$C^{zz} (l,t=0)$ versus spatial separation $l$ (top) and (b) log-log plots of 
equal space-point 
time-dependent correlators $C^{xx}(l=0,t)$ and $C^{zz}(l=0,t)$  (bottom) 
versus $t$. Slopes yield exponents $\chi_x,\chi_z$ and 
$\chi_x/z_x,\chi_z/z_z$, respectively (see text).}\label{loglogx}
\end{center}
\end{figure}

Fluctuations of $\phi_+$ and 
$\phi_-$ propagate with a relative speed between them, thus one can
eliminate the linear propagating term in either (\ref{phi+eq}) or 
(\ref{phi-eq}), but not simultaneously in both. At the linear level, the 
dynamics of $\phi_+$ and $\phi_-$ are mutually decoupled. This implies  
$\chi_+=1/2=\chi_-$ and $z_+=2=z_-$, for the roughness and dynamic exponents 
defined by the correlation functions for $\phi_+$ and $\phi_-$, analogous to (\ref{corrux}-\ref{corruz}).
This implies $\chi_x=1/2=\chi_z$ and $z_x=2=z_z$ in the linear theory. 

With the nonlinear terms, eqs.~(\ref{phi+eq},\ref{phi-eq}), cannot be 
solved exactly and naive perturbative 
expansions in powers of the nonlinear coefficients yield diverging 
corrections in the long wavelength limit. In order to deal with these long 
wavelength divergences in a systematic manner,
we employ perturbative one-loop Wilson momentum shell 
DRG~\cite{Stanley,halpin}. This 
is implemented by first integrating out the 
dynamical fields $\phi_\pm({\bf{ q}},\omega)$ with 
wavevector
$\Lambda/b<q<\Lambda,\,b>1$, perturbatively up to the one-loop order using 
(\ref{phi+eq}-\ref{phi-eq}).$\,\Lambda$ is the wave vector upper cut-off. 
We then 
rescale wavevectors by $q'=bq$, so that the upper cutoff is restored to 
$\Lambda$. The frequency $\omega$ and the fields are also scaled  
appropriately~\cite{Stanley,halpin}.
\begin{figure}[t]
\begin{center}
\includegraphics[scale=0.3,keepaspectratio=true]{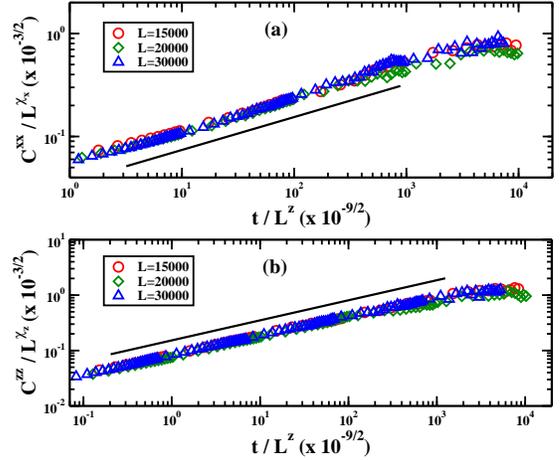}
\caption{(color online). Log-log plots of $C^{xx}$ (top) and $C^{zz}$ (bottom) as 
functions of $t/L^z$, showing data collapse on to single curves after 
scaling.}\label{collapse}
\end{center}
\end{figure}
The one-loop perturbation theory is 
constructed using the bare 
propagators and correlators of $\phi_\pm$. We work in the co-moving frame of 
$\phi_+$ where the 
bare propagators 
(in Fourier space) are of the form 
$G_{0}^{+}(k,\omega)=\frac{1}{D_{+}k^{2}+i\omega}$ 
and $G_{0}^{-}(k,\omega)=\frac{1}{D_{-}k^{2}+i(\omega-\alpha k)}$, 
for $\phi_{+}$ 
and $\phi_{-}$ respectively. 
Thus, at linear order 
$\phi_{+}(k,\omega)=G^{+}(k,\omega)\eta_{+}(k,\omega)$ and 
$\phi_{-}(k,\omega)=G^{-}(k,\omega)\eta_{-}(k,\omega)$. In a similar manner, 
correlators of $\phi_\pm$ in the co-moving 
frame of $\phi_+$ are defined as 
$C_{\phi_{+}\phi_{+}}(k,\omega)=\frac{2A_1}{\omega^{2}+D_{+}^{2}k^{4}}$ and 
$ C_{\phi_{-}\phi_{-}}(k,\omega)=\frac{2A_2}{(\omega-\alpha 
k)^{2}+D_{-}^{2}k^{4}}$. 
Notice that since each of (\ref{phi+eq}, \ref{phi-eq}) can be 
reduced to the standard KPZ equation~\cite{kpz} upon setting appropriate 
coupling constants to zero, the lowest order perturbative corrections to 
$D_\pm,A_1,A_2,a_1$ and $a_2$ 
can clearly be classified into two categories: (i) KPZ-type, which survive in 
the KPZ limit, and (ii) non-KPZ type, which 
vanish in that limit. The KPZ-type diagrams are formally identical to those in the 
pure KPZ problem~\cite{kpz}. The relevant one-loop Feynman diagrams 
are listed in the appendix. 
Retaining only the dominant contributions (all of which arise from the 
respective KPZ-type diagrams), we find the corrections to be
\begin{eqnarray}
\tilde{A}_{1}&= A_{1}\left[1+\frac{a_{1}^{2}A_{1}}{\pi D_{+}^{3}}\int_{\Lambda/b}^{\Lambda} \frac{1}{q^{2}}dq \right],\label{A1} \\
\tilde{A}_{2}&= A_{2}\left[1+\frac{A_{1}^{2}a_{2}^{2}}{\pi A_{2} D_{+}^{3}} 
\int_{\Lambda/b}^{\Lambda} \frac{1}{q^{2}}dq \right],\label{A2}\\
\tilde{D}_{+}&= D_{+}\left[1+\frac{A_{1}a_{1}^{2}}{\pi D_{+}^{3}}\int_{\Lambda/b}^{\Lambda} \frac{k^{2}}{q^{2}}dq \right]  ,\label{D+}\\
\tilde{D}_{-}&= D_{-}\left[1+\frac{A_{1}a_{2}c_{1}}{\pi 2D_{+}^{2}D_{-}}\int_{\Lambda/b}^{\Lambda} \frac{k^{2}}{q^{2}}dq \right] .\label{D-}
\end{eqnarray}
None of  the vertices $a_{1}$, $b_{1}$, $c_{1}$, $a_{2}$, $b_{2}$ and $c_{2}$ 
receive any fluctuation corrections at the one-loop order~\cite{fns}.
Under scalings $x\rightarrow bx$, $t\rightarrow b^{z}t$, 
$\phi_{+}\rightarrow b^{\chi_{+}} \phi_{+}$ and $\phi_{-}\rightarrow 
b^{\chi_{-}}\phi_{-}$, the parameters scale as 
$A_{1}\rightarrow b^{z-1-2\chi_{+}}A_{1}$, $A_{2}\rightarrow b^{z-1-2\chi_{-}}A_{2}$, 
$D_{\pm}\rightarrow b^{z-2}D_{\pm}$. 
On rescaling the momentum cut off and taking the limit $\delta l \rightarrow 0$, 
we get the recursion relations
\begin{eqnarray}
\frac{dD_{+}}{dl} &=& D_{+}[z-2+g]\,, \nonumber \\
\frac{dA_{1}}{dl} &=& A_{1}[z-1-2\chi_{+}+g]\,, \nonumber \\
\frac{dD_{-}}{dl} &=& D_{-}[z-2+\frac{1}{2}mnrg]\,, \nonumber \\
\frac{dA_{2}}{dl} &=& A_{2}[z-1-2\chi_{-}+pn^{2}g]\,, 
\end{eqnarray}
where the coupling constant $g\equiv \frac{A_{1}a_{1}^{2}}{\pi D_{+}^{3}}$ and 
dimensionless constants $m=\frac{D_{+}}{D_{-}}$, 
 $p=\frac{A_{1}}{A_{2}}$, $n=\frac{a_{2}}{a_{1}}$, and $r=\frac{c_{1}}{a_{1}}$.
The renormalized coupling $g$ then obeys
  \begin{equation}
   \frac{dg}{dl}=g[-2g+1]\,,
  \end{equation}
 giving the stable RG fixed point $g^{*}=1/2$. 
The scaling exponents can be extracted from the equations
$\frac{dD_{+}}{dl}=\frac{dA_{1}}{dl}=\frac{dD_{-}}{dl}=\frac{dA_{2}}{dl}=0$ at 
the RG fixed point. This gives $z=3/2$ and $\chi_{+}=\chi_{-}=1/2$, which belong to the 
KPZ universality class. Strong dynamic scaling prevails as the 
dynamic exponents for both the fields $\phi_{+}$ and $\phi_{-}$ are the same. 
Since $u_{x}$ and $u_{z}$ can be written as linear 
 combinations of $\phi_{+}$ and $\phi_{-}$, we have 
$\chi_x=\chi_z=1/2$ and $z_x=z_z=3/2$. The presence of 
propagating modes here is crucial; they render the so-called non-KPZ 
nonlinearities irrelevant in the long wavelength limit so 
the model displays KPZ universality.

Having obtained the scaling exponents in the co-moving frame of $\phi_+$, we now argue that 
the values of these exponents are the same in all reference frames connected by
the Galilean transformation~\cite{fns}. Consider the correlation function, $C_+ (x_1-x_2,t_1-t_2)= 
\langle\phi_+(x_1,t_1)\phi_+(x_2,t_2)\rangle$: under a Galilean transformation, 
$t_{1,2}\rightarrow t_{1,2},\;x_{1,2}\rightarrow x_{1,2} + {\sf{v}} t$, 
where $t$ is the time and ${\sf v}$  the Galilean boost.  
$\,x_1-x_2$ and $t_1-t_2$ are unchanged, hence so is $C_+$. The scaling exponents are thus 
the same in all frames connected by Galilean transformations.

The scaling behavior of the displacement fields $u_x$ and $u_z$ can also be obtained 
numerically by integrating  eqs.~(\ref{dotux},\ref{dotuz2}). The correlation functions 
$C^{xx}(x,t)$ and $C^{zz}(x,t)$ can be calculated from the solutions of these equations. 
The equations of motion for $u_{x}$ and $u_{z}$ are
simulated with diffusion constants $D_{1}\,=\,D_{2}\,=\,1$, wave velocities 
$\lambda_{2} = 0.1$\,, $\lambda_{3} = 0.2$\,,
co-efficients of nonlinear terms $\gamma_{1} = 1.0$\,, $\gamma_{2} = 2.0$ and $\gamma_{3} = 10.0$ and 
time step $dt = 0.01$\,. We simulate a system of $2\times 10^{4}$ particles.
Log-log plots of $C^{xx}(x,0)$ 
and $C^{zz}(x,0)$ are shown in Fig.~\ref{loglogx} (top). We obtain 
$\chi_x=0.47\pm 0.01,\, \chi_z=0.485\pm 0.015$. Similarly, the 
log-log plots of $C^{xx}(0,t)$ and $C^{zz}(0,t)$ shown in Fig.~\ref{loglogx} (bottom) yield $z_x=1.45\pm 
0.05,\,z_z=1.49\pm 0.06$, which are the same as the dynamic exponent for the KPZ universality 
class, within error bars. Our numerical results are thus in close agreement with the DRG 
results. Fig.~\ref{collapse} shows the correlation functions
for different system sizes $L$ collapse on each other on scaling $t$ by $L^{z}$ and 
correlations by 
$L^{\chi_{i}}$, for $i=x, z\,$. This clearly 
establishes universal scaling in the model. Technical details of our 
numerical studies can be found in the SM.

\section{ Conclusions and Outlook} We have shown that a colloidal crystal drifting in an electric field is linearly
stable, with long-wavelength lattice distortions propagating as waves.
For particles of radius $a=1 \mu m$, $\kappa a= 2.5 $, $d=3a$ in an electric field of strength 
$150$ V$/m$, we estimate the speed of the propagating modes to be $~ 10 \mu m/s$.
Using renormalization group methods we establish that, in the drifting steady state,
lattice distortions both transverse and longitudinal to the lattice, display strong dynamic 
scaling with dynamic exponent $3/2$ and belongs to the KPZ universality class. A numerical 
analysis of the equations for the displacement fields confirm these results.

The notion of universality survives even for driven elastic media. 
However, unlike equilibrium, this universal behavior is controlled by 
the drive, displaying 1D KPZ scaling.
While extending our analysis to higher 
dimensions may be nontrivial, we can comment that in higher (D$>$1) dimensions 
there should be 1 longitudinal and D-1 transverse modes. The 
presence of propagating waves should make the system anisotropic. Thus, it 
is unlikely that the fluctuations in higher dimensions belong to the KPZ 
universality class. We look forward to theoretical attempts in understanding the 
universal properties of the fluctuations at higher D and experimental tests of our 
predictions for propagating modes in drifting colloidal cyrstals.

\begin{acknowledgments}
We thank S. Ramaswamy for introducing the problem to us and sharing his insight into it.
AB wishes to thank the Alexander von Humbolt Stiftung (Germany) for partial 
financial support under the Research Group Linkage Programme scheme (2016).
\end{acknowledgments}

\appendix
\begin{widetext}
\section{Equations of motion and diagrammatic expansions}

The equation of motion for $\phi_{\pm}$ are 
\begin{eqnarray}
\dot{\phi}_{+}-\frac{\alpha}{2}\partial_{x}\phi_{+}+a_{1}(\partial_{x}\phi_{+})^
{2}+b_{1}(\partial_{x}\phi_{-})^{2}
+c_{1}(\partial_{x}\phi_{+})(\partial_{x}\phi_{-})
=D_{+}\partial^{2}_{x}\phi_{+}+\eta_{+} \label{phi+eq1}\\
\dot{\phi}_{-}+\frac{\alpha}{2}\partial_{x}\phi_{-}+a_{2}(\partial_{x}\phi_{+})^
{2}+b_{2}(\partial_{x}\phi_{-})^{2}+c_{2}(\partial_{x}\phi_{+})(\partial_{x}\phi_{-})
=D_{-}\partial^{2}_{x}\phi_{-}+\eta_{-}. \label{phi-eq1}
\end{eqnarray}
where the co-efficient of wave term 
$\frac{\alpha}{2}=\sqrt{\lambda_{2}\lambda_{3}}$, noises are 
$\eta_{\pm}=f_{x}\pm \sqrt{\frac{\lambda_{2}}{\lambda_{3}}}f_{z}$ and other 
co-efficients are 
$a_{1}=-\frac{(\gamma_{1}+\gamma_{3})}{4}\sqrt{\frac{\lambda_{3}}{\lambda_{2}}}
-\frac{\gamma_{2}}{4}\sqrt{\frac{\lambda_{2}}{\lambda_{3}}}$, 
$b_{1}=-\frac{(\gamma_{3}-\gamma_{1})}{4}\sqrt{\frac{\lambda_{3}}{\lambda_{2}}}
-\frac{\gamma_{2}}{4}\sqrt{\frac{\lambda_{2}}{\lambda_{3}}}$,
 $c_{1}=\frac{\gamma_{3}}{2}\sqrt{\frac{\lambda_{3}}{\lambda_{2}}}-\frac{\gamma_{2
}}{2}\sqrt{\frac{\lambda_{2}}{\lambda_{3}}}$ 
and $a_{2}=-b_{1}$, $b_{2}=-a_{1}$ and $c_{2}=-c_{1}$. In the special case with 
$\gamma_{1}=2\gamma_{3}$ and $\gamma_{2}/\gamma_{3}=
\lambda_{3}/\lambda_{2}$ Eqs.~(\ref{phi+eq1}-\ref{phi-eq1}) reduce to two 
separate KPZ equations \cite{kpz}.

$ G_{0}^{+}(k,\omega)$ and $G_{0}^{-}(k,\omega)$ are the bare propagators for $\phi_{+}$ 
and $\phi_{-}$ respectively in the co-moving frame of $\phi_+$\, and have the form
\begin{equation}
 G_{0}^{+}(k,\omega)=\frac{1}{D_{+}k^{2}+i\omega} , \hspace{0.5cm} 
 G_{0}^{-}(k,\omega)=\frac{1}{D_{-}k^{2}+i(\omega-\alpha k)}\,.
\end{equation}
The correlators of $\phi_\pm$ in the Fourier space are defined in the co-moving 
frame of $\phi_+$ as 
\begin{equation}
 C_{\phi_{+}\phi_{+}}(k,\omega)=\frac{2A_1}{\omega^{2}+D_{+}^{2}k^{4}}, \hspace{0.5cm}
 C_{\phi_{-}\phi_{-}}(k,\omega)=\frac{2A_2}{(\omega-\alpha 
k)^{2}+D_{-}^{2}k^{4}}.
\end{equation}
Our perturbative Dynamic Renormalization Group (DRG) calculation may be represented diagrammatically \cite{Stanley}. The symbols, that we use are 
explained below.
\begin{figure}[!h]
\begin{center}
\includegraphics[scale=0.6,keepaspectratio=true]{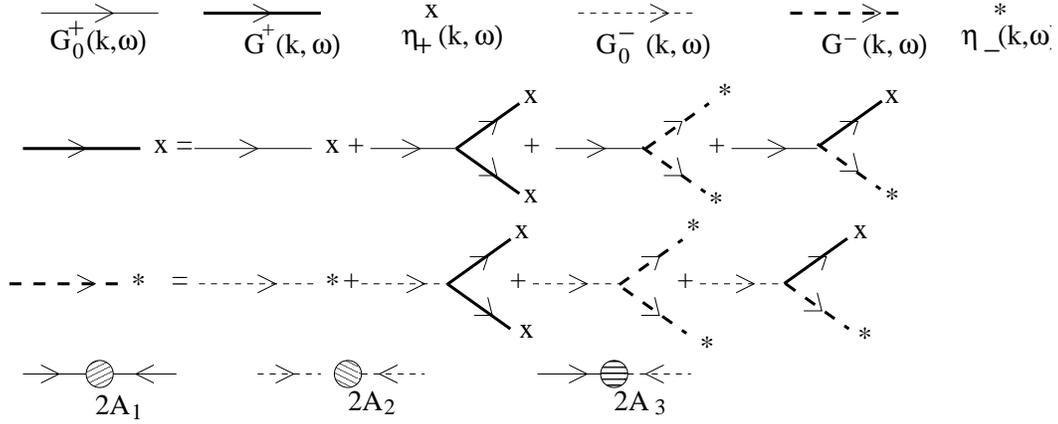}
\caption{Diagrammatic representation of the propagators and noise for $\phi_+$ and $\phi_-$. Perturbation expansion for the propagators
$G_{+}(k, \omega)$, $G_{-}(k, \omega)$. Contracted noise $A_1$, $A_2$, $A_3$. }
\end{center}
\end{figure}

\section{Propagator renormalization}
There are four one-loop diagrams which contribute to the propagator renormalization of $\phi_{\pm}$.
Fig.\,2\, shows the relevant diagrams for propagator renormalization for $\phi_{+}$. 
\begin{figure}[!h]
\begin{center}
\includegraphics[scale=0.65,keepaspectratio=true]{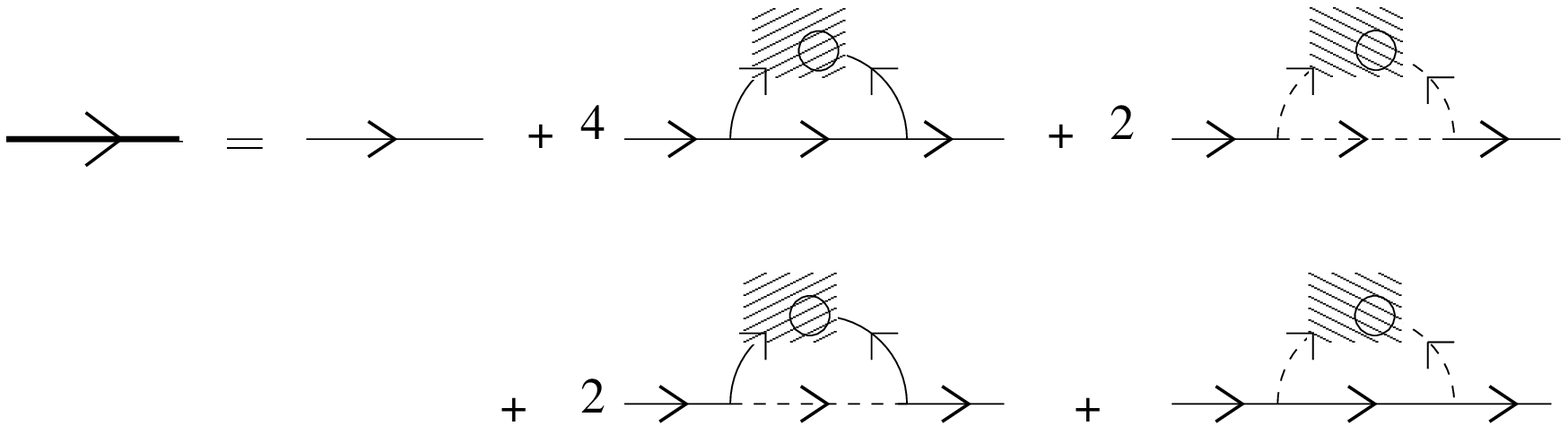}
\caption{One-loop diagrams for propagator renormalization.}
\end{center}
\end{figure}
The renormalized propagator $G^{+}(k,\omega)$ can be written as 
\begin{equation}
 G^{+}(k,\omega)=G_{0}^{+}(k,\omega)+T_{1}+T_{2}+T_{3}+T_{4}
\end{equation}
where $T_{1}$ and $T_{2}$ contain contributions only from $\phi_+$ and 
$\phi_-$, respectively, and $T_{3}$, $T_{4}$ are the contributions from both 
the fields.
We calculate the individual contributions below. 
\begin{equation}
\begin{multlined}
T_{1}=\frac{8A_{1}a_{1}^{2}G_{0}^{+}(k,\omega)^{2}}{(2\pi)^{2}}\int dq\int_{-\infty}^{\infty} d\Omega [q(k-q)][-qk]
G_{0}^{+}(q,\Omega)G_{0}^{+}(-q,-\Omega)G_{0}^{+}(k-q,\omega-\Omega) \\
=\frac{8A_{1}a_{1}^{2}G_{0}^{+}(k,\omega)^{2}}{(2\pi)^{2}}\int dq\int_{-\infty}^{\infty} d\Omega \frac{[q(k-q)][-q k]}
{[D_{+}q^{2}+i\Omega][D_{+}q^{2}-i\Omega][D_{+}(k-q)^{2}-i\Omega]}\\
\end{multlined}
\end{equation}
After angular integration $T_{1}$ behaves as $\sim \int dq \frac{\pi 
k(k-q)}{D_{+}^{2}[k^{2}-2kq+2q^{2}]}
\approx k^{2}\int \frac{dq}{k^{2}-2kq+2q^{2}} \sim k^{2}I_{a}$ where 
$I_{a}\sim\frac{1}{k}$. This is the contribution that survives in the KPZ limit 
of the model equations. 
Next we calculate the contribution coming from the second diagram which scales 
as 
\begin{equation}
 \begin{multlined}
  T_{2}\approx \int dq\int_{-\infty}^{\infty} d\Omega [q(k-q)][-qk]G_{0}^{-}(q,\Omega)G_{0}^{-}(-q,-\Omega)G_{0}^{-}(k-q,\omega-\Omega) \\
  \approx \int dq\int_{-\infty}^{\infty} \frac{d\Omega [q(k-q)][-qk]}{[D_{-}q^{2}+i\Omega-i\alpha q][D_{-}q^{2}-i\Omega+i\alpha q]
  [D_{-}(k-q)^{2}-i\Omega-i\alpha(k-q)]}\\
  \approx \int dq \frac{\pi k(k-q)}{D_{-}[i\alpha 
k-D_{-}(k^{2}-2kq+2q^{2})]}\approx k^{2}I_{b},
 \end{multlined}
\end{equation}
where $I_{b}\sim\frac{1}{\sqrt{k}}$. Thus, $T_{1}$ is more divergent compared 
to $T_{2}$. Third diagram has a contribution 
\begin{equation}
 \begin{multlined}
  T_{3}\approx \int dq\int_{-\infty}^{\infty} d\Omega \frac{[q(k-q)][-qk]}{[D_{+}q^{2}+i\Omega][D_{+}q^{2}-i\Omega][D_{-}(k-q)^{2}-i\Omega-i\alpha(k-q)]}\\
  \approx \int \frac{dq}{D_{-}q^{2}-D_{-}(k-q)^{2}+i\alpha(k-q)}\sim k^{2}I_{c}
 \end{multlined}
\end{equation}
where $I_{c}\sim -\ln k$ and the contribution from the last diagram is $T_{4}\sim -k^{2}\int \frac{dq}{[i\alpha q-D_{-}q^{2}-D_{+}(k-q)^{2}]}=k^{2}I_{d}$ with 
$I_{d}\sim-\ln k^{2}$\,. So, $T_{1}$ is the most relevant diagram which gives the renormalized propagator for $\phi_{+}$ of the form
\begin{equation}
 G^{+}=G_{0}^{+}-\frac{A_{1}a_{1}^{2}G_{0}^{+}(k,\omega)^{2}}{D_{+}^{2}\pi}\int_{\Lambda/b}^{\Lambda} \frac{k^{2}}{q^{2}}dq\,.
\end{equation}

Similarly, we find the renormalized propagator for $\phi_{-}$ and will be of the form 
\begin{equation}
G^{-}=G_{0}^{-}-\frac{A_{1}a_{2}c_{1}G_{0}^{-}(k,\omega)^{2}}{2D_{+}^{2}\pi}
\int_{\Lambda/b}^{\Lambda} \frac{k^{2}}{q^{2}}dq.
\end{equation}
Notice that in the hydrodynamic limit $k\rightarrow 0$, the dominant 
corrections to both
$G_0^+$ and $G_0^-$ are from the contributions that survive in the KPZ limit of 
the 
model.
From the corrections to $G_0^+$ and $G_0^-$, we obtain the fluctuation 
corrected diffusion constants:
\begin{eqnarray}
  \tilde{D}_{+}=D_{+}\left[1+\frac{A_{1}a_{1}^{2}}{\pi D_{+}^{3}}\int_{\Lambda/b}^{\Lambda} \frac{k^{2}}{q^{2}}dq \right]\hspace*{0.5cm}
  \nonumber \\
  \tilde{D}_{-}=D_{-}\left[1+\frac{A_{1}a_{2}c_{1}}{2\pi D_{+}^{2}D_{-}}\int_{\Lambda/b}^{\Lambda} \frac{k^{2}}{q^{2}}dq\right] 
\end{eqnarray}
\section{Noise renormalization}
Consider now the  noise renormalization for 
$\phi_{+}$ field. The corresponding one-loop corrections receive 
contributions  from one diagram that survives in the KPZ limit and one that 
vanishes in the KPZ limit. 
The diagrammatic representation of the perturbation series for the noise renormalization of $A_{1}$ is shown in figure below.

\begin{figure}[!h]
\begin{center}
\hspace{-3cm}
\includegraphics[scale=0.5,keepaspectratio=true]{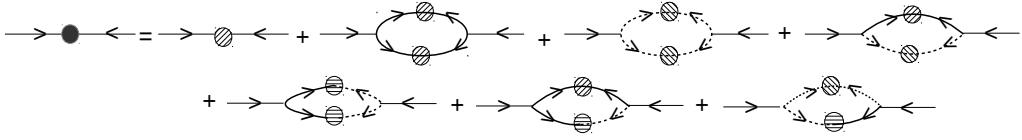}
\caption{One-loop diagrams for the renormalization of $A_1$.}\label{noisefig}
\end{center}
\end{figure}
The first one is 
given by
\begin{equation}
\begin{multlined}
 I_{1}\sim \int_{\Lambda/b}^{\Lambda} dq \int_{-\infty}^{\infty} d\Omega \frac{q^{2}(k-q)^{2}}{[\Omega^{2}+D_{+}^{2}q^{4}][\Omega^{2}+D_{+}^{2}(k-q)^{4}]} \\
  \sim \int_{\Lambda/b}^{\Lambda} dq \frac{1}{k^{2}+2q^{2}-2kq} \sim \int_{\Lambda/b}^{\Lambda} \frac{dq}{q^{2}+k^{2}/2}
 \end{multlined}
\end{equation}
that survives in the KPZ limit of the model. The additional contribution 
that vanishes in that limit  is 
\begin{equation}
\begin{multlined}
 I_{2} \sim \int_{\Lambda/b}^{\Lambda} dq  \int_{-\infty}^{\infty}d\Omega \frac{q^{2}(k-q)^{2}}{[(\Omega-\alpha q)^{2}+D_{-}^{2}q^{4}]
 [(\Omega+\alpha(k-q))^{2}+D_{-}^{2}(k-q)^{4}]} \\
 \sim \int_{\Lambda/b}^{\Lambda} dq 
\frac{k^{2}-2kq+2q^{2}}{D_{-}k^{2}\alpha^{2}+(k^{2}-2kq+2q^{2})^{2}D_{-}^{3}}. \\
 \end{multlined}
\end{equation}
The first contribution, $I_1$ is the dominant contribution in the 
thermodynamic limit $k^2\rightarrow 0$ and $I_2$ is 
subleading. This may be understood as follows. Notice that the most significant 
(or the dominant) contribution to both $I_1$ and $I_2$ from the lower (i.e., 
small-$q$) limits of the integrals, which are controlled by $k^2$.
Set $q=0$ in both the integrands in $I_1$ and $I_2$: The 
respective integrands scale as $\sim \frac{1}{k^{2}}$ and  $\sim 
\frac{k^{2}}{k^{2}+k^{4}}$. For  small enough $k^2$,  $k^2\gg k^{4}$, 
yielding $I_1\gg I_2$ in the limit $k\rightarrow 0$, establishing the dominance 
of $I_1$ over $I_2$ in the limit $k\rightarrow 0$.

There are four more diagrams (see Fig.~\ref{noisefig}) for noise correlations 
whose contributions are 
clearly subdominant to the contribution from $I_1$ above. Thus, in the long 
wavelength limit, $I_1$, the contribution that is nonvanishing in the KPZ limit 
of the model, determines the fluctuation correction to $A_1$. We then have
\begin{equation}
 \tilde{A_{1}}=A_{1}+\frac{a_{1}^{2}A_{1}^{2}}{\pi D_{+}^{3}}\int_{\Lambda/b}^{\Lambda} \frac{1}{q^{2}}dq ,\label{A1}
\end{equation}
Similarly, renormalized $A_{2}$ will be 
\begin{equation}
 \tilde{A_{2}}=A_{2}+\frac{A_{1}^{2}a_{2}^{2}}{\pi D_{+}^{3}} 
\int_{\Lambda/b}^{\Lambda} \frac{1}{q^{2}}dq. \label{A2}
\end{equation}
Again, the dominant contribution in the hydrodynamic limit is the contribution 
that survives in the KPZ limit of the model.

\section{Vertex renormalization}
The diagrams that contribute to the vertex renormalization for $a_{1}$ are shown 
Fig.~\ref{afig}. 
\begin{figure}[!h]
\begin{center}
\includegraphics[scale=0.6,keepaspectratio=true]{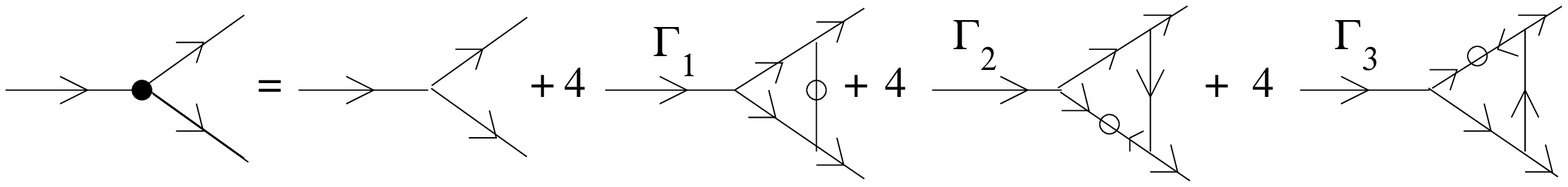}
\caption{One-loop diagrams that contribute to the fluctuation-corrections to 
the vertex $a_{1}$.}\label{afig}
\end{center}
\end{figure}
Renormalized vertex $\tilde{a}_{1}=a_{1}(1+\Gamma_{1}+\Gamma_{2}+\Gamma_{3})$ 
where $\Gamma_{1}$,  $\Gamma_{2}$ and $\Gamma_{3}$ are 
three different vertices as shown in the figure where  
$\Gamma_{1}=\frac{2a_{1}^{2}A_{1}}{\pi 
D_{+}^{3}}\int_{\Lambda/b}^{\Lambda}\frac{dq}{q^{2}}$ 
and $\Gamma_{2}=\Gamma_{3}=-\frac{a_{1}^{2}A_{1}}{\pi D_{+}^{3}}\int_{\Lambda/b}^{\Lambda}\frac{dq}{q^{2}}$.
So $\tilde{a}_{1}=a_{1}$. There are similar relevant diagrams for $b_{1}$ renormalization which also give $\tilde{b}_{1}=b_{1}$.
Similarly, it can be shown that all the vertices $a_{1}$, $b_{1}$, $c_{1}$, 
$a_{2}$, $b_{2}$ and $c_{2}$ receive no fluctuation corrections that diverge in 
the limit $k\rightarrow 0$. We discard all the finite corrections in the spirit 
of DRG calculations.

\section{Flow equations}
Of the total momentum range $0<|q|<\Lambda$, the high momenta components $\Lambda e^{-\delta l}<|q|<\Lambda$
 are integrated out and we rescale in such a way so that the momentum cut off remains same. Taking the limit 
 $\delta l \rightarrow 0$, we get the recursion relations
\begin{eqnarray}
\frac{dD_{+}}{dl} = D_{+}[z-2+g] \hspace*{1.5cm}\nonumber \\
\frac{dA_{1}}{dl} = A_{1}[z-1-2\chi_{+}+g] \hspace*{0.6cm}\nonumber \\
\frac{dD_{-}}{dl} = D_{-}[z-2+\frac{1}{2}mnrg] \hspace*{0.5cm}\nonumber \\
\frac{dA_{2}}{dl} = A_{2}[z-1-2\chi_{-}+pn^{2}g]  
\end{eqnarray}
 where the coupling constant $g\equiv \frac{A_{1}a_{1}^{2}}{\pi D_{+}^{3}}$ and some dimensionless constants are 
 $m=\frac{D_{+}}{D_{-}}$, $p=\frac{A_{1}}{A_{2}}$, $n=\frac{a_{2}}{a_{1}}$, and $r=\frac{c_{1}}{a_{1}}$. The coupling 
 constant has a flow equation $\frac{dg}{dl}=g[-2g+1]$ which gives the stable RG fixed point $g^{*}=1/2$. Those 
 dimensionless constants $m$, $p$, $n$ and $r$ have the flow equations 
$\frac{dm}{dl}= m[1-\frac{1}{2}nrm]g$, $ \frac{dp}{dl} = p[2(\chi_{-}-\chi_{+})+(1-np^{2})g]$, 
$ \frac{dn}{dl} = n(\chi_{+}-\chi_{-})$ and 
$\frac{dr}{dl} = r(\chi_{-}-\chi_{+})$\,. Under the scale transformations 
$x\rightarrow bx$, $t\rightarrow b^{z}t$, $\phi_{+}\rightarrow b^{\chi_{+}} 
\phi_{+}$ and  $\phi_{-}\rightarrow b^{\chi_{-}}\phi_{-}$\,.
To get the fixed points we should set the LHS of the flow equations equal to zero. Flow equations of $m$, $p$, $n$ and $r$ give 
$n^{*}r^{*}m^{*}=2$, $p^{*2}n^{*}=1$ and $\chi_{+}=\chi_{-}$. We use
 these relations and put $\frac{dD_{+}}{dl}=\frac{dA_{1}}{dl}=\frac{dD_{-}}{dl}=\frac{dA_{2}}{dl}=0$ which give the exponents $z=3/2$ and 
 $\chi_{+}=\chi_{-}=1/2$, which belong to the KPZ universality class.

\section{Numerical simulation}
We numerically integrate Eqs.~(2-3) in the main text, calculate the 
time-dependent correlation functions of $u_x$ and $u_z$, which yield the 
scaling exponents in the hydrodynamic limit, and compare with the DRG results. 
The discretized equations used for numerical simulation are as follows:
\begin{eqnarray}
u_{x}(x,t+\Delta t)= u_{x}(x,t)+\frac{\lambda_{2}}{2}\left[u_{z}(x+1,t)-u_{z}(x-1,t)\right]dt+ 
\frac{\gamma_{1}}{4}\left[u_{x}(x+1,t)-u_{x}(x-1,t)\right] \hspace*{1.5cm}\nonumber \\
\left[u_{z}(x+1,t)-u_{z}(x-1,t)\right]dt
+D_{1}\left[u_{x}(x+1,t)-2u_{x}(x,t)+u_{x}(x-1,t)\right]dt+\sqrt{2N_{1}dt}\,\zeta_{1}(x,t)  \\
u_{z}(x,t+\Delta t)= 
u_{z}(x,t)+\frac{\lambda_{3}}{2}\left[u_{x}(x+1,t)-u_{x}(x-1,t)\right]dt+\frac{\gamma_{2}}{
4}\left[u_{x}(x+1,t)-u_{x}(x-1,t)\right]^{2}dt \hspace*{1.1cm}\nonumber \\
+\frac{\gamma_{3}}{4}[u_{z}(x+1,t)-u_{z}(x-1,t)]^{2}dt+D_{2}[u_{z}(x+1,t)-2u_{z}
 (x,t)+u_{z}(x-1,t)]dt +\sqrt{2N_{2}dt}\,\zeta_{2}(x,t) 
\end{eqnarray}
$\zeta_{1}$ and $\zeta_{2}$ are Gaussian random variables with zero mean and variances 
$\sqrt{2N_{1}dt}, \sqrt{2N_{2}dt}$ respectively. In the simulation, random initial conditions were used with 
periodic boundary conditions.

Roughness exponents are defined by the spatial scaling of the equal-time 
correlators $C^{xx}(l,0)\sim l^{\chi_{x}}$ and 
$C^{zz}(l,0)\sim l^{\chi_{z}}$ where $l=|x'-x|$. Growth exponent is defined through the correlation function with a time delay 
$ C^{xx}(0,t)\sim t^{\beta_{x}}$ and $C^{zz}(0,t)\sim t^{\beta_{z}}$\,. 
These two exponents define dynamic exponents: $z_{x}=\chi_{x}/\beta_{x}$ and 
$z_{z}=\chi_{z}/\beta_{z}$. These correlation functions are shown in Fig.~2 
of the main text.


\end{widetext}
\bibliography{EP3.bib}
\end{document}